\documentclass[pre,aps,twocolumn,superscriptaddress]{revtex4-2}
\usepackage{standalone}
\usepackage{graphicx,color}

\input{scrload}
\usepackage{amsmath,bm,url}
\usepackage{hyperref}
\usepackage{tikz}
\definecolor{liquid}{rgb}{0.67, 0.8, 0.94}
\definecolor{slab}{rgb}{0.5, 0.5, 0.5}
\definecolor{grad}{rgb}{0.0, 0.28, 0.67}
\definecolor{border}{rgb}{0.4,0.4,0.4}

\newcommand{\br}{{\bm r}}

\newcommand{\bu}{{\bm u}}

\newcommand{\bs}{{\bm s}}

\newcommand{\bj}{{\bm j}}

\newcommand{\bq}{{\bm q}}
\newcommand{\bp}{{\bm p}}

\newcommand{\rme}{{\rm e}}
\newcommand{\kb}{{k_\textsc{B}}}
\newcommand{\LE}{{\textsc{le}}}

\newcommand{\ud}{{\mathrm d}}
\newcommand{\bla}{{\big\langle}}
\newcommand{\bra}{{\big\rangle}}

\input{scrload}

\begin{document}

\title{Fluid flow at interfaces driven by thermal gradients}

\author{Pietro Anzini}
\affiliation{Dipartimento di Scienza e Alta Tecnologia, Universit\`a degli Studi dell’Insubria, Via Valleggio 11, 22100 Como, Italy}
\affiliation{To.Sca.Lab, Universit\`a degli Studi dell’Insubria, Via Valleggio 11, 22100 Como, Italy}

\author{Zeno Filiberti}
\affiliation{Dipartimento di Scienza e Alta Tecnologia, Universit\`a degli Studi dell’Insubria, Via Valleggio 11, 22100 Como, Italy}

\author{Alberto Parola}
\email[]{alberto.parola@uninsubria.it}
\affiliation{Dipartimento di Scienza e Alta Tecnologia, Universit\`a degli Studi dell’Insubria, Via Valleggio 11, 22100 Como, Italy}

\begin{abstract}
Thermal forces drive several nonequilibrium phenomena able to set a fluid in motion without pressure gradients. Although the
most celebrated effect is thermophoresis, also known as \hbox{Ludwig-Soret} effect, probably the simplest example where thermal 
forces are at play is \hbox{thermo-osmosis}: The motion of a {\it confined} fluid exclusively due to the presence of a temperature gradient. 
We present a concise but complete derivation of the microscopic theory of \hbox{thermo-osmosis} based on linear response theory.
This approach is applied to a simple fluid confined in a slab geometry, mimicking the flow through a pore in a membrane 
separating two fluid reservoirs at different temperatures. We consider both 
the case of an open channel, where the fluid can flow freely, and that of 
a closed channel, where mass transport is inhibited and a pressure drop sets in at the boundaries. 
Quantitative results require the evaluation of generalized transport coefficients, but a preliminary check on a specific 
prediction of the theory has been successfully performed via nonequilibrium molecular dynamics simulations. 
\end{abstract}
\maketitle

\section{Introduction}
In a bulk fluid at constant pressure a thermal gradient
can not exert a net force on the fluid particles~\cite{landau_fluid}: 
Fluid motion in homogeneous systems can only be induced by
external forces, such as gravity or pressure gradients.
However, in the presence of a confining surface (or, more generally, 
in an inhomogeneous environment), a fluid flow develops due to the thermal gradient.
This effect, now referred to as {\it thermo-osmosis}~\footnote{Feddersen, who discovered this phenomenon,
gave it the name of {\it thermodiffusion}~\cite{feddersen_1972}. The same phenomenon in gases is also known as {\it thermal creep}~\cite{kennard_1938kinetic}
or {\it thermal transpiration} (Reynolds~\cite{reynolds_1879}).
The term \hbox{thermo-osmosis} (or thermal osmosis) was introduced by Lippmann\cite{lippmann_1907} 
in the liquid regime.}, was observed for the first time by
Feddersen~\cite{feddersen_1972} in 1873, who measured the 
\hbox{temperature-induced} motion of air 
through a tube fitted with porous plugs of gypsum or spongy platinum. The gas
drift was directed towards the warmer side as long as a temperature difference
between the sides of the porous partition was present. 
More quantitative investigations of \hbox{thermo-osmosis} in gases have been indirectly spurred by
the invention of the radiometer by Crookes~\cite{maxwell_reynolds_radiometer}. 
The purpose of the radiometer~\footnote{The radiometer (or
light mill) is a small glass bulb, partially evacuated, inside of which
a set of vanes is accommodated on a spike. Each vane is shiny on one side
and blackened on the other to reflect and adsorb the incident light.} 
was to detect the pressure of light~\cite{crookes_repulsion}. However, 
as shown by the work of Maxwell~\cite{maxwell_1879}, Schuster~\cite{Schuster_exp} and Reynolds~\cite{reynolds_1879},
its motion is due to the \hbox{thermo-osmotic} flow which develops
near the edges of the vanes~\cite{kennard_1938kinetic,Piazza2020}, 
and not to the momentum transfer due to the incident electromagnetic radiation.
More recently, Sone and Yoshimoto proposed a simple experiment~\cite{sone_exp}
to demonstrate the onset of \hbox{thermo-osmosis}, showing that, at sufficiently low pressure,
the \hbox{thermo-osmotic} flow which develops near the surface can even overwhelm convection.

The same effect also occurs in liquids, but the magnitude of the flow turns
out to be much smaller than in gases. This is probably the reason why 
it took many years after the discovery of Feddersen before that Lippman was able 
to detect the \hbox{thermo-osmotic} flow  
of water through a membrane of gelatin separating two volumes held at different temperatures~\cite{lippmann_1907}.
A few years later Aubert~\cite{aubert_1912} addressed the problem more systematically and
found that, when subject to a temperature difference, some membranes originate a water flow from the cold to the hot region
whereas other in the opposite direction.
\hbox{Thermo-osmosis} in liquids was rediscovered in the forties by
Derjaguin and Sidorenkov~\cite{derjaguin_sidorenkov_1941thermoosmosis}, who were 
not aware of the works of Lippman and Aubert. 
The group at the Russian Academy of Sciences studied \hbox{thermo-osmosis} across different membranes 
and capillaries~\cite{surface_forces_1987}, but, as understood later, their
results were strongly influenced by the presence of free charges
in the membrane~\cite{hutchisonnixondenbigh_1948}.
A further complication is that the direction of the \hbox{thermo-osmotic} flow 
can change depending on the temperature, as shown by Haase and de Greiff, who studied \hbox{thermo-osmosis} of 
water through a cellophane membrane at different temperatures~\cite{HaasedeGreiff_1965} and 
reported an inversion of the effect at temperatures higher than $60^\circ\,{\rm C}$.

Currently thermal osmosis is an accepted phenomenon and a renewed interest is stimulated
in relation to possible applications to fuel cells, water management, desalination 
and water recovery~\cite{ESSALHI2021279,barragankjelstrup_2016review}. 

Since the studies by Derjaguin, many authors have measured the pressure gradient induced by \hbox{thermo-osmosis}
through membranes and capillaries under different conditions,
but the experimental results often disagree about the direction and the magnitude 
of \hbox{thermo-osmotic} fluxes: The apparently simple phenomenon of thermal osmosis 
is not yet fully characterised (and understood) at a microscopic level.
For a review see Ref.s~\cite{shukla_review_1984,barragankjelstrup_2016review} and references therein.
A recent work, claiming the first microscale observation of the velocity
field imposed by \hbox{thermo-osmosis}, goes towards this direction~\cite{bregulla_2016},
but the results seem to be affected by the presence of surface charge.

Actually, one of the reasons at the origin of the contradictory results found 
in the liquid regime is the lack of a deep understanding of the phenomenon
through a microscopic theory, able to account for the perturbation on the liquid 
structure (and dynamics) in a few molecular layers near the wall. 
\hbox{Thermo-osmosis} in rarefied gases~\footnote{In the limit where the mean free path
is larger than the range of the interparticle potential.} is to date accurately
predicted by kinetic theories~\cite{kennard_1938kinetic,sone_2000}.  
Maxwell in 1879 obtained the expression (see Eq.~(\ref{eq:maxwell_v}) for the \hbox{thermo-osmotic} velocity 
of a gas subject to a thermal gradient parallel to a confining surface~\cite{maxwell_1879}.
His derivation unravels the mechanism behind \hbox{thermo-osmosis} in gases, 
namely the longitudinal transfer of momentum during the collision between 
the particle and the surface~\cite{kennard_1938kinetic}.
\newline
In the opposite limit, the liquid regime, \hbox{thermo-osmosis} is described in 
the language of (macroscopic) irreversible thermodynamics~\cite{de_ma_1984}. 
This approach was proposed many years ago by 
Derjaguin~\cite{derjaguin_sidorenkov_1941thermoosmosis,hutchisonnixondenbigh_1948,surface_forces_1987} and identifies the driving force as 
the local enthalpy change induced by the confining surface.
At the moment, most of the numerical and experimental works on \hbox{thermo-osmosis} in liquids 
essentially rely on this theory for the interpretation of their results~\cite{barragankjelstrup_2016review,joly18,jivkov_21,joly22}.
In particular, in molecular dynamics simulations the velocity profile is obtained 
by evaluating the excess enthalphy near the wall, which acts as the force term 
in the linearized \hbox{Navier-Stokes} equations.
However, the hypothesis underlying continuum theories is that the relevant 
observables vary on a length scale much larger than the typical range of the
interaction: 
Near a surface this condition is no longer satisfied because the fluid properties
eventually driving the phenomenon may display strong, 
but short-ranged, modulations. In addition, the viscosity, which is assumed to 
be constant in the whole system~\cite{surface_forces_1987,joly18,jivkov_21,joly22}, is perturbed near the the interface~\cite{todd_PRE_07,todd_PRL_08,evans_morriss_2008,galliero_2012}. 
For this reason, the accuracy of the results based on the classical macroscopic paradigm are still under debate.
Recently, a series of simulations on a model system~\cite{ganti17,ganti18,proesmans_19}, where spurious effects 
due to the charge and exotic confining potentials are not present, 
helped to gain a deeper understanding of the origin of \hbox{thermo-osmosis} in liquids.
The main focus of the simulations was the direct measure of the ``thermal force" acting on the fluid particles, due to the presence of a thermal gradient. The results were then compared with available expressions, coming either from a ``mechanical route" or nonequilibrium thermodynamics, showing that only the latter approach is able to provide a good agreement with the numerical simulations. 

Prompted by these studies, a microscopic derivation of the \hbox{thermo-osmotic} flow based on linear response theory was developed~\cite{prl_2019}.
The first formulation of the theory focused on \hbox{thermo-osmosis} in the 
simplest configuration, an infinite open channel (slit) 
without boundaries at its ends (see Fig.~\ref{channel}).
In this work we provide a critical derivation of the theory and
we extend our approach to a closed channel (slit), a geometry 
particularly relevant for experiments in membranes 
and simulations in systems without boundary conditions.
We also deduce the equations in the case of cylindrical 
geometry, relevant for the description of the flow in pores and nanotubes.
Moreover, a novel physical interpretation of the equations is presented,
demonstrating a close correspondence between our approach, 
based on \hbox{Kubo-Mori} formalism, and the phenomenological expressions derived 
by Derjaguin~\cite{surface_forces_1987}.
This result sheds light on the interpretation of the numerical 
simulations performed in Ref.s~\cite{ganti17,ganti18}.
To test our model, we derived analytically the scaling of the pressure drop for a large channel and we verified 
these results through molecular dynamics simulations in a closed \hbox{two-dimensional} system.

The paper is organised as follows.
In Section~\ref{micro} we provide a complete, critical derivation 
of the theoretical framework. In Section~\ref{channel} we derive specific predictions in both open and closed channel geometry.  
The results from the numerical simulations are presented in Section~\ref{sim}. 

\section{Microscopic Theory}
\label{micro}

\subsection{The Model}
Although the microscopic approach described in this Section applies to general Hamiltonian systems, we will 
concentrate on the most popular model of simple fluid~\cite{simpleliquids_fourth}: A collection of $N$ classical particles 
mutually interacting via the spherically symmetric pair potential $v(|\bq_i-\bq_j|)$, possibly 
under the effects of an external field $V(\bq)$ mimicking the presence of confining walls.
The  microscopic Hamiltonian density of such a system can be written as
\begin{align}
\hat {\cal H}(\br) &= \sum_i \delta(\bq_i-\br)\,\hat h_i \nonumber \\
&= \sum_i \delta(\bq_i-\br) \Biggl[ \frac{p_i^2}{2m} + V(\bq_i) + \frac{1}{2} \sum_{j (\ne i)} v(q_{ij}) \Biggr],  \nonumber 
\end{align}
where the shorthand notation $q_{ij} = |\bq_i-\bq_j|$ has been introduced.
We note that this definition suffers from some arbitrariness, due to the \hbox{non-local} nature of the interparticle 
potential $v(q_{ij})$: In this expression, the energy of the pair $(i,j)$ is attributed half to each particle~\cite{rowlinson_1993}. 
Here and in the following, ``hat denotes a function defined in the $6N$ dimensional phase space of the system.
The Hamiltonian of the model is then given by the integrated Hamiltonian density: 
\begin{equation}
\hat H = \int \ud\br\, \hat {\cal H}(\br).
\notag
\end{equation}

The Liouville operator, acting on a function $\hat A$ defined in the phase space, is written in terms of the 
Poisson brackets as ${\scr L} = \{ \hat H,\cdot \}$ and governs the time evolution of every observable:
\begin{equation*}
\frac{\ud \hat A}{\ud t} = \frac{\partial \hat A}{\partial t} -{\scr L} \hat A. 
\end{equation*}
Another important quantity is the phase space distribution function 
$\hat F(t)$ which provides the probability of a given microscopic configuration $(\bq_1\cdots\bq_N,\bp_1\cdots\bp_N)$.
As such, it satisfies the normalization condition $\int \ud\Omega\, \hat F(t) =1$, where $\ud\Omega$ is the volume element in the phase space. 
In systems out of equilibrium,
the phase space distribution $\hat F(t)$ is generally a function of time, but in a steady state $\hat F$ becomes
time independent. The time evolution of the phase space distribution is again expressed in terms of the Liouville operator as
\begin{equation}
\frac{\ud \hat F}{\ud t} =  \frac{\partial \hat F}{\partial t} + {\scr L} \hat F.
\label{eq:evo_f}
\end{equation} 
In full thermodynamic equilibrium, the distribution $\hat F$ is 
given by the Boltzmann expression $\hat F_0 = Z^{-1} \,\rme^{-\beta \hat H}$ in terms of the inverse temperature $\beta=(\kb T)^{-1}$. 
The normalization factor $Z$ is the canonical partition function. Being $\hat H$ independent of time, the phase distribution $\hat F_0$
is indeed a stationary solution of the evolution equation. 

\subsection{Linear Response Theory}

Our goal is to describe the steady state of a possibly inhomogeneous fluid in a temperature gradient induced by 
two different temperatures at the opposite boundaries of the sample.
Because of the ensuing \hbox{non-uniform} temperature, standard equilibrium statistical mechanics cannot be straightforwardly adopted. 
Even the natural concept of Local Equilibrium (LE), a condition where the basic relations among thermodynamic bulk quantities hold also locally,
is just a first approximation to the actual phase space distribution~\cite{balescu}. This can be proved by 
first defining the most general energy density function of the model
\begin{equation}
\hat {\cal E}(\br) = \hat {\cal H}(\br) -\bu(\br)\cdot\hat \bj(\br)
-\mu(\br)\hat \rho(\br)
\label{energy}
\end{equation}
in terms of the previously defined Hamiltonian density $\hat {\cal H}(\br)$ and of the densities of the other microscopic conserved quantities, namely
the momentum density
\begin{equation}
\hat j^{\alpha}(\br) = 
\sum_i p_i^{\alpha} \,\delta(\br - \bq_i)
\label{eq:mic_mom_curr}
\end{equation}
and the mass density 
\begin{equation}
\hat \rho(\br) = m\,\sum_i \delta(\br - \bq_i).
\notag
\end{equation}
Here $\bu(\br)$ and $\mu(\br)$ are the fields related to the velocity profile and the \hbox{space-dependent} chemical potential (per unit mass): External 
parameters identifying the local velocity and chemical potential of the LE state. These two fields, together 
with the field related to the inverse temperature profile $\beta(\br)$, define the most general LE distribution function:
\begin{equation}
\hat F_\LE = Z_\LE^{-1} \,\exp\left [ - \int \ud\br \,\beta(\br)\,\hat {\cal E}(\br) \right ].
\label{fle}
\end{equation}
In the special case of uniform external fields $\beta,\bu,\mu$ this distribution does indeed 
describe the equilibrium state of our system: A fluid flowing at uniform velocity $\bu$. 
Therefore, for slowly varying fields, it is natural to expect that this LE distribution will provide a 
faithful description of the state of the system. However, in a fluid close to a confining surface, all physical 
properties, and then also the external fields $\beta(\br),\bu(\br),\mu(\br)$, vary considerably on the scale 
of the correlation length making the LE assumption questionable. 

It is well known that $\hat F_\LE$ is not a stationary solution of the evolution equation~(\ref{eq:evo_f}), as shown by the explicit evaluation of its time derivative: 
\begin{eqnarray}
\frac{\ud \hat F_\LE}{\ud t} &=&  {\scr L} \hat F_\LE =  \hat F_\LE\, 
\int \, \ud\br\,\bigg\{ -\beta\,\partial_\alpha\hat J^\alpha_H \nonumber \\
&& + \beta\,u^\alpha\, \left [ \partial_\nu\,\hat J_j^{\alpha\nu} 
- \hat \rho\,\partial_\alpha V \right ]
+ \beta \,\mu\,\partial_\alpha\,\hat j^\alpha \bigg\},  
\label{evol}
\end{eqnarray}
where the dependence on the local position $\br$ is understood. Here, $\partial_\alpha$ is the
partial derivative with respect to $r^\alpha$ and the summation over repeated Greek indices is implied. 
Mass, energy and momentum currents $\hat j^\alpha(\br)$, $\hat J^\alpha_H$, $\hat J_j^{\alpha\nu}$ 
are defined by the Poisson brackets:
\begin{align}
\label{massc}
{\scr L} \hat \rho(\br) &= \left\{\hat H, \hat \rho(\br)\right\} = \partial_\alpha\hat j^\alpha(\br);  \\
\label{enerc}
{\scr L} \hat {\cal H}(\br) &= \left\{\hat H, \hat {\cal H}(\br)\right\} = \partial_\alpha\hat J_H^\alpha(\br);  \\
\label{momc}
{\scr L} \hat j^\alpha(\br) &= \left\{\hat H, \hat j^\alpha (\br)\right\} 
=\partial_\nu\hat J_j^{\alpha\nu}(\br)  + \frac{\hat\rho(\br)}{m}\partial_\alpha V(\br). 
\end{align}
The mass current actually coincides with the previously defined momentum density $\hat \bj(\br)$. Instead, the 
energy and momentum currents are not uniquely defined by Eqs.~(\ref{enerc},\ref{momc}) providing 
only their divergence. This ambiguity originates from 
the presence of \hbox{non-local} terms in the pair interaction contribution, as 
thoroughly discussed in the literature~\cite{henderson,rowlinson_1993}. The explicit expressions for the energy 
and momentum currents are written as: 
\begin{eqnarray}
\label{energyc}
\hat J_H^{\alpha}(\br) &=& \sum_i \frac{p_i^\nu}{m} \,\left  [ \hat h_i\,\delta(\br-\bq_i) \delta^{\alpha\nu} + 
\Gamma_i^{\alpha\nu}(\br)\right ]; \\
\hat J_j^{\alpha\nu}(\br) &=& \sum_i \left [ \frac{p_i^\alpha p_i^\nu}{m} \,\delta(\br-\bq_i) + \Gamma_i^{\alpha\nu}(\br)\right ],
\end{eqnarray}
where the \hbox{non-local} contribution is defined by 
\begin{equation}
\Gamma_i^{\alpha\nu}(\br) = 
\frac{1}{2} \sum_{j(\ne i)} \frac{\partial v(q_{ij})}{\partial q_i^\alpha}\int_{C_{ij}}{\hskip -0.3cm \ud s^\nu} \,\delta(\br-\bs) 
\label{prob}
\end{equation}
and depends on the (arbitrary) choice of the path $C_{ij}$ connecting the position $\bq_i$ of particle $i$ to the position $\bq_j$ 
of particle $j$~\cite{henderson}. Despite this intrinsic ambiguity in the definition of the energy and momentum currents, we stress
that the expression~(\ref{evol}) is well defined, depending uniquely on the divergence of the currents. 

Equation~(\ref{evol}) shows unambiguously that $\hat F_\LE$  evolves in time and then it 
cannot represent a stationary phase space distribution: The 
properties of the stationary state of a fluid in a thermal gradient cannot be simply evaluated according to the LE hypothesis~\cite{balescu,zubarev1974nonequilibrium}. 
In order to introduce the correction terms, we follow the classical treatment by Mori under the assumption that 
the deviations from LE are small, i.e. that the effects of the external perturbations keeping the systems out of equilibrium 
can be taken into account to linear order (linear response theory)~\cite{mori56,mori58,balescu}. Starting at time $t=0$ with a LE phase 
distribution, after a long time, the stationary state can be formally defined, without approximation, by the distribution
\begin{equation}
\hat F = \hat F_\LE + \lim_{\tau\to\infty} \int_0^\tau \ud t \, \rme^{{\scr L} t} \, {\scr L} \hat F_\LE.
\label{steady}
\end{equation}
The limit $\tau\to\infty$ requires some care. To be well defined it has to be performed at the end of the averaging process
because only averaged quantities possess a definite limit at long times, while in an isolated system,
the phase space distribution itself evolves according to the Hamiltonian dynamics~(\ref{eq:evo_f}). 
The key quantity ${\scr L} \hat F_\LE$ has been evaluated in Eq.~(\ref{evol}). By performing an integration by 
parts we obtain
\begin{eqnarray}
{\scr L} \hat F_\LE &\sim& 
\hat F_0\int \ud\br\,\Big\{ \hat J^\alpha_H \, \partial_\alpha\beta
- \hat J_j^{\alpha\nu} \partial_\nu \left [ \beta\,u^\alpha\right ]  \nonumber \\
&&\qquad \quad- \hat \rho\,\partial_\alpha V\,\beta \,u^\alpha
- \hat j^\alpha\partial_\alpha \left [ \beta \,\mu\right ] \Big\},  
\label{lfe}
\end{eqnarray}
where the assumption of small deviations from equilibrium has been enforced by substituting the equilibrium
distribution function $\hat F_0$ in place of $\hat F_\LE$ at right-hand side in Eq.~(\ref{evol}).
Now, by use of Eqs.~(\ref{steady},\ref{lfe}) we can evaluate the average of any observable in the stationary state. In 
particular, Eq.~(\ref{steady}) shows that 
the LE result has to be corrected with the contribution coming from the time evolution of the phase space distribution. 

As a first step we evaluate the LE averages of the relevant quantities previously defined. A straightforward calculation gives,
to first order in the deviations from thermodynamic equilibrium, all the 
relevant observables:
\begin{itemize}
\item mass density:
\begin{equation}
\bla\hat \rho(\br)\bra_\LE = \rho_0(\br)\Big\vert_{\beta(\br),\mu(\br)},
\label{mle}
\end{equation}
\item momentum density (or mass current):
\begin{equation}
\bla\hat j^\alpha(\br)\bra_\LE = \rho_0(\br) \,u^\alpha(\br),
\end{equation}
\item energy current: 
\begin{equation}
\bla\hat J_H^\alpha(\br)\bra_\LE = \beta \,\int \ud\br^\prime \,
\bla\hat J_H^\alpha(\br)\hat j^\nu(\br^\prime)\bra_0\, u^\nu(\br^\prime),
\end{equation}
\item momentum current:
\begin{align}
\bla\hat J_j^{\alpha\nu}(\br)\bra_\LE = p_0^{\alpha\nu}(\br)
- \int \ud &\br^\prime \Big\langle \hat J_j^{\alpha\nu}(\br)\Bigl[\hat \Delta(\br^\prime)
  \nonumber \\
& - \bla\hat \Delta(\br^\prime)\bra_0 
\Bigr]\Big \rangle_0,
\label{ple}
\end{align}
\end{itemize}
where $\rho_0(\br)$ and $p_0^{\alpha\nu}(\br)$ are the mass density and the pressure 
tensor at equilibrium, evaluated at the average temperature and chemical potential, and
\begin{equation}
\hat \Delta(\br) = \big[ \beta(\br)-\beta\big] \,\hat {\cal H}(\br)
- \big[ \beta(\br)\mu(\br)-\beta\mu\big] \,\hat\rho(\br).
\notag
\end{equation}
One might expect that both the mass density and the momentum current in LE would coincide with their 
equilibrium expressions evaluated at the local temperature and chemical potential. While this expectation 
is correct for the mass density~(\ref{mle}) and the diagonal components of the momentum current, in general Eq.~(\ref{ple}) 
allows for nonvanishing off-diagonal components of $\langle\hat J_j^{\alpha\nu}(\br)\rangle_\LE$. 

Analogously, by use of Eqs.~(\ref{steady},\ref{lfe}), we can evaluate the corrections to the LE averages, but we will not report
the general, rather lengthy, expressions because the chosen geometry will considerably simplify the results. 

\subsection{Steady State}

Within Mori's formalism, linear response theory provides corrections to a known LE phase space distribution $\hat F_\LE$. 
This implies that we need to know the external \hbox{space-dependent} fields $\beta(\br), \bu(\br), \mu(\br)$ defining $\hat F_\LE$
via~(\ref{energy},\ref{fle}). However, in a real experimental set-up this is not the case: We can certainly tune the physical parameters,  
like temperature, at the boundaries of the system, but the actual temperature profile in the bulk of the fluid is self-consistently
determined, if the approach to equilibrium is governed by the system's Hamiltonian. 

In order to determine the external field we need five additional equations defining the steady state. The most natural procedure 
is to impose the vanishing of the time derivative of the averaged densities $\langle\hat\rho(\br)\rangle$, $\langle\hat {\cal H}(\br)\rangle$ and 
$\bla\hat \bj(\br)\bra$ which satisfy the appropriate continuity equations:
\begin{eqnarray}
\label{cm}
\partial_t \bla \hat\rho(\br)\bra &+& \partial_\alpha \bla\hat j^\alpha(\br)\bra = 0; \\
\label{ch}
\partial_t \bla\hat{\cal H}(\br)\bra &+& \partial_\nu \bla\hat J_H^\nu(\br)\bra = 0; \\
\label{cj}
\partial_t \bla\hat j^\alpha(\br)\bra &+& \partial_\nu \bla\hat J_j^{\alpha\nu}(\br)\bra +
\frac{\langle\hat\rho(\br)\rangle}{m} \partial_\alpha V(\br) = 0,
\end{eqnarray}
where the averages are taken with the full steady state phase space distribution~(\ref{steady}).
In this way, we have five new equations enforcing the stationarity of the state. Written in terms of
the spatial divergence of the previously defined currents, these equations are going to identify the consistent
stationary profile of temperature, velocity, and chemical potential. 

\section{Thermo-osmosis in a channel}
\label{channel}
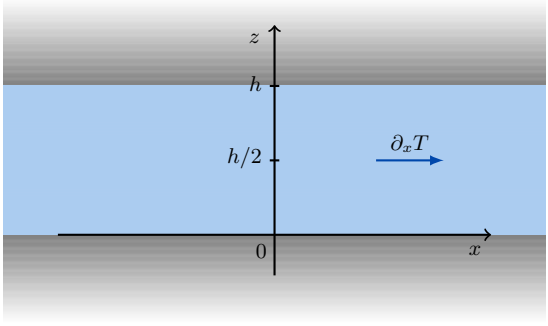
\begin{figure}
\centering
\begin{tikzpicture}[scale=0.9,every node/.style={scale=0.9}]

\shade[top color=slab,bottom color =white] (0,-0.5) rectangle (8,0.8) ;
\shade[top color=white,bottom color =slab] (0,4.3) rectangle (8,3) ;
\shade[top color=liquid,bottom color =liquid] (0,3) rectangle (8,0.8) ;

\draw[thick,->] (0.8,0.8) -- (7.2,0.8) node[below right,xshift=-0.45cm, yshift=-1pt] {$x$};
\draw[thick,->] (4,0.2) -- (4,3.9) node[below right,xshift=-0.5cm, yshift=0pt] {$z$};

\draw (4,0.8) node[anchor=east,yshift=-7pt] {$0$};
\draw (4,3) node[anchor=east,xshift=-2pt,yshift=1pt] {$h$};
\draw (4,1.9) node[anchor=east,xshift=-2pt,yshift=1pt] {$h/2$};
\draw[thick] (4cm+2pt,3cm) -- (4cm-2pt,3cm) node[anchor=east] {};
\draw[thick] (4cm+2pt,1.9cm) -- (4cm-2pt,1.9cm) node[anchor=east] {};

\draw[thick,->,arrows={-latex},grad] (5.5,1.9) -- (6.5,1.9) node[above,xshift=-0.5cm, yshift=0pt] {};
\draw[thick] (6,1.9) node[above,xshift=0cm, yshift=0pt] {$\partial_x T$};

\end{tikzpicture}
\caption{Schematic representation of the slab geometry. The $y$ direction is perpendicular to the plane of the sheet.
}
\label{fig:geoslab}

\end{figure}

Now we apply the previously outlined strategy to the simplest geometry supporting a three dimensional flow: An 
infinite slab in the $(x,y)$ plane confined by a potential $V(z)$ in the $z$-direction. The thermal gradient is set in 
the $x$ direction by suitably choosing the boundary conditions at $x\to\pm\infty$, 
i.e. keeping the two boundaries at $x\to\pm\infty$ at different temperatures, 
uniform in the $(y,z)$ plane. Therefore it is natural to expect that
the system keeps uniform along $y$: 
Both temperature and chemical potential vary linearly along the $x$ direction, 
while the velocity field $\bu(\br)$ is directed along $x$ and changes with $z$. 
On this basis, we look for 
a solution to the continuity equations~(\ref{cm},\ref{ch},\ref{cj}) of the form 
\begin{align}
\label{beta}
\beta(\br) &=  \beta + x\,\partial_x\beta; \\
\label{mu}
\beta(\br)\mu(\br) &= \beta\mu + x\,\partial_x [\beta\mu]; \\
\label{u}
u^\alpha(\br) &= \delta^{\alpha x}\,u^x(z),
\end{align}
where $\beta=(\kb T)^{-1}$ and $\mu$ correspond to the average value of the inverse temperature and the chemical potential respectively and $\partial_x\beta$, $\partial_x [\beta\mu]$ are space independent.
It is convenient to introduce the ratio between these
two gradients because it is going to play a role in the definition of the velocity profile:
\begin{equation}
\gamma = \frac{\partial_x [\beta\mu]}{\partial_x\beta}.
\label{gamma}
\end{equation}

First, we have to evaluate the steady state averages of the mass density and mass, energy, and momentum currents. 
By exploiting the symmetry properties of the chosen geometry, several terms disappear. Both static and dynamic equilibrium
correlation functions odd in $x$ or $y$ must vanish by symmetry after spatial integration, while equilibrium correlation functions odd in $z$ are allowed by the 
slab geometry. The mass density therefore reduces to its LE expression~(\ref{mle}), while the non vanishing components of the currents become:
\begin{itemize}
\item mass current:
\begin{eqnarray}
\bla\hat j^x(z)\bra &=& \rho_0(z)\,u^x(z) + 
\int_0^\infty \ud t \int \ud\br^\prime  \nonumber \\
&&\Big \{ \bla \hat j^x(\br,t)\,
\hat J^x_Q(\br^\prime)
\bra_0\,\partial_x\beta
\nonumber \\
&& - \beta\,\bla \hat j^x(\br,t)\,\hat J_j^{xz}(\br^\prime)\bra_0 \,\partial_{z^\prime}u^x(z^\prime) 
\Big \},
\label{massj}
\end{eqnarray}
\item energy current: 
\begin{eqnarray}
\bla\hat J_H^x(z)\bra &=& \beta \int \ud\br^\prime \bla\hat J_H^x(\br)\hat j^x(\br^\prime)\bra_0 \,u^x(z^\prime)
\nonumber \\
&&
+\int_0^\infty \ud t \int \ud\br^\prime  
\Big \{\bla \hat J_H^x(\br,t)\,
\hat J^x_Q(\br^\prime)
\bra_0\,\partial_x\beta
\nonumber \\
&& -\, \beta\,\bla \hat J_H^x(\br,t)\,\hat J_j^{xz}(\br^\prime)\bra_0 \,\partial_{z^\prime}u^x(z^\prime) 
\Big \},
\end{eqnarray}
\item diagonal momentum current:
\begin{equation}
\bla\hat J_j^{\alpha\alpha}(z)\bra = p_0^{\alpha\alpha}(z)\Big\vert_{\beta(x),\mu(x)}
\label{stressd}
\end{equation}
($\alpha$ is not summed),
\item off-diagonal momentum current:
\begin{eqnarray}
\bla\hat J_j^{xz}(z)\bra &=& -\int \ud\br^\prime \,x^\prime\,\bla \hat J_j^{xz}(\br) \,
\hat {\scr P}(\br^\prime)
\bra_0\,\partial_x\beta;
\nonumber \\
&& + \int_0^\infty \ud t \int \ud\br^\prime  
\Big \{ \left \langle \hat J_j^{xz}(\br,t)\,
\hat J^x_Q(\br^\prime)
\right \rangle_0\,\partial_x\beta \nonumber \\
&& - \beta\,\left \langle \hat J_j^{xz}(\br,t)\,\hat J_j^{xz}(\br^\prime) \right \rangle_0\,\partial_{z^\prime} u^x(z^\prime)
\Big \}.
\label{stress}
\end{eqnarray}
\end{itemize}
Here we have defined the ``heat current" operator as 
\begin{equation}
\hat J^x_Q(\br) = \hat J^x_H(\br)- \gamma\, \hat j^x(\br),
\label{heat}
\end{equation}
together with the additional conjugate operator
\begin{equation}
\hat {\scr P}(\br) = \hat {\cal H}(\br)- \gamma\, \hat \rho(\br).
\label{p}
\end{equation}
We stress that, contrary to bulk fluids, in inhomogeneous systems the odd-rank correlation functions do not necessarily vanish
because isotropy is broken. 
The explicit expressions for the currents immediately show that the two continuity equations $\partial_\alpha \bla \hat j^\alpha(\br)\bra = 0$ and 
$\partial_\alpha \bla \hat J_H^\alpha(\br)\bra = 0$ are identically satisfied by our Ansatz: Only the $x$ component
of the mass and energy currents does not vanish, but their statistical averages do not depend on $x$. The only \hbox{non-trivial} continuity 
equations are those involving the momentum current. The continuity equation for the $z$ component of the momentum density is 
\begin{equation}
\partial_\alpha \bla \hat J_j^{z\alpha}(\br)\bra = -\frac{\langle\hat\rho(\br)\rangle}{m}\,\frac{\ud V(z)}{\ud z} 
\notag
\end{equation}
and reduces to the hydrostatic equilibrium condition 
\begin{equation}
\frac{\ud p_0^{zz}(z)}{\ud z}\Bigg\vert_{\beta(x),\mu(x)} = -\frac{\rho_0(z)}{m}\Bigg\vert_{\beta(x),\mu(x)}\,\frac{\ud V(z)}{\ud z},
\notag
\end{equation}
which is identically satisfied by the equilibrium normal pressure at any temperature and chemical potential.
Therefore, the remaining continuity equation for the $x$ component of the momentum density 
$\partial_\alpha \langle \hat J_j^{x\alpha}(\br)\rangle = 0$ contains the only relevant information on the structure of the velocity profile: 
\begin{eqnarray}
&&\partial_x p_0^{xx}(z)\Big\vert_{\beta(x)\mu(x)} -\partial_z\int \ud\br^\prime \,x^\prime\,\bla \hat J_j^{xz}(\br) \,
\hat {\scr P}(\br^\prime) \bra_0\,\partial_x\beta 
\nonumber \\
&&\quad + \,\partial_z\int_0^\infty \ud t \int \ud\br^\prime  
\Big \{ \bla \hat J_j^{xz}(\br,t)\,
\hat J^x_Q(\br^\prime)
\bra_0\,\partial_x\beta \nonumber \\
&&\quad - \,\beta\bla \hat J_j^{xz}(\br,t)\,\hat J_j^{xz}(\br^\prime) 
\bra_0\,\partial_{z^\prime} u^x(z^\prime)
\Big \} = 0.
\label{pxx}
\end{eqnarray}
Equation~(\ref{pxx}) allows to find the velocity field $u^x(z)$ in the fluid, showing the validity of our Ansatz.
The equation can be concisely written in the form
\begin{equation}
\label{gold}
\int_0^{h} \ud z^\prime {\cal K}(z,z^\prime)\,  \partial_{z^\prime}u^x(z^\prime) =
\partial_x\beta \, \big[ {\cal S}_s(z)+{\cal S}_d(z)\big],
\end{equation}
where the kernel ${\cal K}(z,z^\prime)$ has the physical meaning of local viscosity
\begin{equation}
{\cal K}(z,z^\prime) = \beta \int_0^\infty \ud t^\prime \int \ud\br^\prime_\perp    \bla\hat J_j^{xz}(\br,t^\prime)\hat J_j^{xz}(\br^\prime)\bra_0.  
\label{cappa}
\end{equation}
On right-hand side, ${\cal S}_s(z)$ and ${\cal S}_d(z)$ represent the static and dynamic source terms
\begin{eqnarray}
{\cal S}_s(z) &=&
-\int_z^{\frac{h}{2}} \ud z^\prime \,
\big[ \partial_\beta + \gamma\, \partial_{\beta\mu}\big]  p_0^{xx}(z^\prime)
\nonumber\\
&& -\int \ud\br^\prime \,x^\prime\,\bla \hat J_j^{xz}(\br) \,
\hat {\scr P}(\br^\prime) \bra_0;
\label{ss}
\\
{\cal S}_d(z) &=&
\int_0^\infty \ud t \int \ud\br^\prime  \bla \hat J_j^{xz}(\br,t)\,
\hat J^x_Q(\br^\prime) \bra_0,
\label{sd}
\end{eqnarray}
where we have chosen the integration constant in the static source term so to preserve the symmetry of the problem
upon reflection across the middle of the slab
and the equilibrium tangential pressure $p_0^{xx}(z)$ is taken as a function of the 
independent thermodynamic variables $\beta$ and $\beta\mu$.
Note that all averages appearing in Eqs.~(\ref{cappa},\ref{ss},\ref{sd}) are performed in thermal equilibrium.

Summarizing: The Ansatz~(\ref{beta},\ref{mu},\ref{u}) provides a consistent solution of the continuity equations
in the stationary state. This means that the thermal gradient $\partial_x\beta$ is fully determined by the boundary 
conditions at $x=\pm\infty$, while the gradient of the velocity field is the solution of the integral equation 
(\ref{gold}). The only unknown parameters are the ratio $\gamma$~(\ref{gamma}) defining the chemical potential gradient 
and an undetermined constant shift in the velocity field, coming from the solution of Eq.~(\ref{gold}). Both these apparent arbitrariness have a deep physical meaning 
that will be discussed in the next Sections. 

\subsection{The static source term}
\label{sst}

The previously derived expression of the static source term~(\ref{ss}) contains the thermodynamic derivatives 
of the transverse component of the 
pressure tensor plus a static correlation function, whose physical meaning is not transparent. Moreover, both terms are 
\hbox{ill-defined} because of the known ambiguities in the definition of the momentum current $\hat J^{\alpha\beta}_j(\br)$ 
in inhomogeneous environments~\cite{henderson,rowlinson_1993}. 
However, it can be shown that {\it the sum} of these two terms, hence the full static 
source term, is a \hbox{well-defined} quantity. 
This result follows from the identity (see Appendix~\ref{sec:sst_divergence}):
\begin{equation}
\partial_z\,{\cal S}_s(z) = -\int \ud\br^\prime \,x^\prime \,\partial_\alpha
\bla\hat J_j^{x\alpha}(\br)\,\Delta\hat{\scr P}(\br^\prime)\bra_0.
\label{source1}
\end{equation}
with  $\Delta\hat{\scr P}(\br^\prime) = \hat{\scr P}(\br^\prime) - \big<\hat{\scr P}(\br^\prime) \big>_0$.
The divergence of the momentum current, appearing in Eq.~(\ref{source1}), 
is a \hbox{well-defined} quantity, as discussed in Ref.~\cite{henderson}, 
showing that the static source term is indeed unambiguously defined. 
Similar arguments allow to show that also the dynamic source term is well defined, as expected.

We can also write Eq.~(\ref{source1}) in an alternative, more transparent way (see Appendix~\ref{sec:sst_divergence}):
\begin{equation}
\partial_z\,{\cal S}_s(z) = - \kb T \big[ h^v_0(z) - \gamma\,\rho_0(z) \big], 
\label{final}
\end{equation}
where 
\begin{align}
\label{hmicro}
\quad &h_0^v(z)=\left [ \frac{5}{2}\kb T +V(z)\right ] n_0(z)  \\
&+\frac{1}{2} \int \ud\br^\prime \, n_2(\br,\br^\prime)
\left [ v(s) -  \frac{\ud v(s)}{\ud s} \,\frac{(x-x^\prime)^2}{s}\right ]_{s=|\br-\br^\prime|} \nonumber 
\end{align}
is the {\it transverse} component of the {\it virial} enthalpy density in thermal equilibrium. 

Notice that in inhomogeneous environments the enthalpy density is indeed a tensor: In planar geometry, ``transverse'' indicates the $xx$ (or $yy$) component of the enthalpy tensor. As in the case of all thermodynamic observables which depend on the inter-particle potential, the enthalpy density can not be defined without ambiguities in \hbox{non-homogeneous} systems. 
Our derivation only shows that this specific expression for the enthalpy, {\it i.e.} the {\it virial} one, 
is mathematically equivalent to the well-defined form~(\ref{source1})
of the static source term. However, it would be misleading to infer that this is the ``correct" definition
of the enthalpy density close to a wall: As shown in Ref.~\cite{rowlinson_1993}, no unique definition for the enthalpy density can be found in inhomogeneous environments.

Integrating in $z$, and using the symmetry about the middle of the channel, we find
\begin{equation}
{\cal S}_s(z) = \kb T \int_z^{\frac{h}{2}} \ud z^\prime\, \big[ h^v_0(z^\prime) - \gamma\,\rho_0(z^\prime) \big].
\label{ss2}
\end{equation}
This simple formula is exact within Linear Response Theory for a slab geometry and, as it will become clear in Sect.~\ref{sec:open_channel}, is closely related to the 
result for the slip velocity originally obtained by Derjaguin~\cite{surface_forces_1987}  by use of nonequilibrium thermodynamics. 
The static source term originates from the spatial dependence of the (transverse) fluid enthalpy close to a 
confining surface. Note, however, that linear response theory predicts the presence of an additional ``dynamic" source 
term~(\ref{sd}) and includes in the kernel~(\ref{cappa}) also the effects of spatial inhomogeneities on the fluid viscosity.

In this Section we refer to a simple slab geometry, but the results can be easily generalized to a more physical cylindrical (pore) geometry, as outlined in Appendix~\ref{pore}.

\subsection{Open Channel}
\label{sec:open_channel}

We previously mentioned that our equations allow to determine the velocity profile subject to the definition of two unknown constants: A uniform shift in the velocity field and the ratio $\gamma$~(\ref{gamma}). 
It is not surprising that Eq.~(\ref{gold}) provides the {\it gradient} of the velocity field $u^x(z)$: The
fluid Hamiltonian  is invariant under a Galileo transform in the $x$ direction and then the absolute fluid velocity is determined by the boundary condition at the wall surface alone. 

Regarding the appropriate value of $\gamma$, we observe that 
in order to unambiguously determine the solution, we must supplement our equations with
additional conditions on the physical properties of the system at \hbox{$x\to\pm\infty$}.
In the geometry we dub ``open channel", 
if we want to let the fluid flow freely in the $x$ direction, it is natural to impose the vanishing of the pressure gradient along $x$, at 
least far from the confining surfaces. In this way, the fluid motion will be clearly attributed to thermal, rather than mechanical, forces. 
Taking the channel width $h$ sufficiently large, we assume that the fluid in the central region of the channel ($z\sim\frac{h}{2}$) does not 
feel the effects of the confining surfaces. Therefore, in that region, the pressure tensor, which coincides with the LE result
(\ref{stressd}), is isotropic and given by the bulk value $p_0$. The vanishing of the gradient along $x$ then implies that
\begin{equation}
\partial_\beta p_0 + \gamma\,\partial_{\beta\mu} p_0 = 0.
\notag
\end{equation} 
Using standard thermodynamic relations, 
this equation allows to identify the unknown ratio $\gamma$ which equals the bulk enthalpy per unit mass at equilibrium:
\begin{equation}
\gamma = \frac{\partial_x [\beta\mu]}{\partial_x \beta} = -\frac{\partial_\beta p_0}{\partial_{\beta\mu} p_0} = h_m.
\label{open}
\end{equation} 
Moreover, with this choice, the definition of the static source term~(\ref{ss}) simplifies because
\begin{equation}
\big[ \partial_\beta + \gamma\, \partial_{\beta\mu}\big] \,p_0^{xx}(z) = \partial_\beta p^{xx}_0(z)\Big\vert_{p_0},
\notag
\end{equation}
where the derivative is taken at constant bulk pressure. Analogously, also the alternate expression~(\ref{final}) of the static term simply becomes
\begin{equation}
{\cal S}_s(z) = \kb T \int_z^{\frac{h}{2}} \ud z^\prime\, \Delta h^v_0(z^\prime), 
\notag
\end{equation}
where $\Delta h^v_0(z^\prime)$ is just the difference between the local virial enthalpy density and $h_m\,\rho(z)$, which represents the enthalpy density in the local density approximation.

Having fixed the value of the unknown constant $\gamma$, we can now solve Eq.~(\ref{gold}) for the gradient of the velocity field 
$\partial_z u^x(z)$. Unfortunately, this would require the evaluation of several dynamical correlation functions. 
Therefore, an exact expression for the velocity profile 
can be found only at low densities, i.e., in the ideal gas limit.
In the opposite limit, namely in liquids, the 
dynamical correlations can not be evaluated analytically, 
and a simplified form can be justified only in a region sufficiently far from the confining walls.

\subsubsection*{Liquid phase}
The kernel ${\cal K}(z,z^\prime)$ depends on both coordinates $z$ and $z^\prime$ because of the broken 
translational invariance in the $z$ direction. Sufficiently far from the confining surfaces, it becomes a function of the 
single combination $\zeta=z-z^\prime$. Moreover, its integral $\int \ud\zeta\,{\cal K}(\zeta)$ coincides with the shear viscosity $\eta$
defined through the standard Green-Kubo formula~\cite{balescu}. Then, the integral kernel of Eq.~(\ref{gold}) can be physically 
interpreted as a local viscosity and, far from the confining surfaces, it is expected to decay in $\zeta$. 
Far from the walls, for a slowly varying velocity field $u^x(z)$, the left 
hand side of Eq.~(\ref{gold}) can then be identified as $\eta\,\partial_z u^x(z)$. 

Moreover, the source terms at right-hand side of Eq.~(\ref{gold}) are 
\hbox{non-vanishing} only in a neighborhood of the confining surfaces. With the choice~(\ref{open}) of $\gamma$, the static source term~(\ref{ss2}) 
is different from zero only where the wall affects the equilibrium property of the fluid, i.e. within 
a distance of the order of the correlation length or of the range of the wall-particle interaction.
The dynamic source term~(\ref{sd}) 
vanishes in a bulk fluid, being the integral of an odd-rank tensor dynamical correlation function, and is expected to be different from zero
only within a few correlation lengths or a few mean free paths from the walls. 
Therefore, the gradient of the velocity field, obtained by solving 
Eq.~(\ref{gold}), will decay to zero on some microscopic length scale from the wall. 
This means that the velocity profile of the fluid, obtained from 
the expression of the mass current~(\ref{massj}), will tend to a constant far from the surface.

This asymptotic value is precisely the ``creep 
velocity" introduced by Maxwell~\cite{kennard_1938kinetic}. 
However, we want to stress that in our model such a slip velocity is actually undetermined. 
The constant pressure boundary condition allows to fix only one of the two unknown constants, 
$\gamma$, while the other, i.e. the shift in the velocity field, cannot be established as long as we 
do not break the Galileo invariance of the model.
For instance, surface roughness might provide \hbox{no-slip} boundary conditions at the walls, setting the absolute 
velocity at $z=0$ and $z=h$ to zero and then fixing the asymptotic fluid velocity far from the walls.

In a liquid, the particle mean free path is small and then 
we expect that the velocity profile will be mainly determined by the static source term, which embodies the 
strong local deviation of the fluid properties induced by the confining wall. In this case, and adopting the 
previously introduced approximate form of the kernel~(\ref{cappa}), we get the equation for the velocity field:
\begin{equation}
\eta\,\partial_z u^x(z) = - \int_z^{\frac{h}{2}} \ud z^\prime\, \Delta h^0_v(z^\prime) \,\frac{\partial_x T}{T},
\notag
\end{equation}
whose solution, imposing no-slip boundary conditions at the wall surface and in the limit of a wide channel, is 
\begin{equation}
u^x(z) = - \int_0^{\infty} \ud z^\prime\, {\rm Min}\,(z,z^\prime)\,\Delta h^v_0(z^\prime) \,
\frac{\partial_x T}{\eta T}.
\notag
\end{equation}
With these approximations, the slip velocity therefore acquires precisely the form predicted by Derjaguin:
\begin{equation}
v_s = \lim_{z\to\infty} u^x(z)= - \int_0^{\infty} \ud z^\prime\, \,z^\prime\,\Delta h^v_0(z^\prime) \,\frac{\partial_x T}{\eta T}.
\label{slip}
\end{equation}
Note that, within linear response theory, the velocity field $u^x(z)$ coincides 
with the physical velocity only 
far from the surface, where the additional contributions to the mass flux~(\ref{massj}) vanish. 

Having established the equations governing thermo-osmosis (within linear response theory) 
we can revisit the numerical results obtained in Ref.s~\cite{ganti17,ganti18} 
and the problems they raised. 
In these studies a numerical evaluation of the stress induced by a thermal
gradient on a fluid confined in a slab was attempted. Three possible definitions 
of the stress were considered: Two based on the direct evaluation of the 
gradient of the tangential component of the pressure tensor via the {\it Irving-Kirkwood} and 
the {\it virial} expression, and and one based on irreversible  thermodynamics. 
The latter formula linked the stress to the excess enthalpy in the 
boundary layer close to the confining surface, and 
led to a fluid slip velocity in agreement with Derjaguin results. 
In Ref.~\cite{ganti17} the  three routes were found to give different 
results, while the direct numerical evaluation of the thermal force was obtained in Ref.~\cite{ganti18}
showed that only the expression based on irreversible 
thermodynamics was able to reproduce the simulation results. 
The interpretation of these findings directly follows from linear response theory:
\begin{itemize}
    \item[-] The stress acting on the fluid element, expressed by the static source 
    term~(\ref{ss}), does indeed contain the temperature derivative of the 
    tangential component of the pressure tensor but an additional contribution is also present.
    This further term is essential to remove the intrinsic ambiguity
    in the definition of the pressure tensor, as discussed in Section \ref{sst}. 
    Clearly, keeping only one of the two terms and evaluating the pressure according to one of the (infinite) possible definitions of the pressure tensor~\cite{henderson} leads to inconsistent results. 
    \newline
    In addition, in inhomogeneous environments the virial definition of the stress tensor does not correspond to any contour in Eq.~(\ref{prob}) and, more importantly, does not fulfill the hydrostatic balance condition~\cite{henderson}. 
    The virial expression is not an allowed choice for the pressure tensor, although when the system is homogeneous and isotropic Eq.~(\ref{prob}) becomes path independent and the pressure reduces to the virial expression~\cite{henderson}. 
    Therefore any result obtained by applying the virial definition of the stress tensor 
    turns out to be inconsistent (also within our exact linear response approach).
    \item[-] In Section~\ref{sst} it has been shown that the full static source term
    can be written in terms of the excess enthalpy~(\ref{ss2}). The resulting 
    microscopic expression of the excess enthalpy~(\ref{hmicro}) 
    precisely  reproduces the virial route adopted in Ref.~\cite{ganti18}.  
    On this basis, the good agreement with the simulations is expected.  
    \item[-] When we supplement this expression of the static source term with 
    the additional approximation of constant viscosity near the surface, 
    we recover the result by Derjaguin for the velocity slip~(\ref{slip}). 
    \item[-] However, we must add that in this macroscopic interpretation 
    of thermo-osmosis the effect of the dynamic source term~(\ref{sd}) has been neglected, as well as the terms expressing the \hbox{non-trivial} relation between the
    velocity field $u^x(z)$ and the mass current  $\bla \hat j^x(z)\bra$ in~(\ref{massj}).
    As a consequence, we expect that the 
    macroscopic expression based on Derjaguin's result will fail at low density, 
    where the  relevant length scale in thermo-osmosis is given by the mean free 
    path. 
\end{itemize}

\subsubsection*{Gas phase}
Equations~(\ref{gold}) and~(\ref{massj}) are exact to linear order 
in the perturbing fields. 
In the low-density phase, where the interparticle interactions can be 
neglected, Eq.~(\ref{gold}) can be solved analytically and the evaluation of 
Eq.~(\ref{massj}) far from the walls provides the same analytical 
expression for the fluid flow obtained in~\cite{maxwell_1879}.

As stressed in the introduction, thermo-osmosis in gases is guided by 
the specificity of the gas-surface interaction: As realized by Maxwell~\cite{maxwell_1879},
in the case of a perfectly reflecting hard wall, without any momentum
exchange between the particles and the surface, no flow can occur.
Therefore, to obtain a \hbox{non-vanishing} flow, we must introduce in the model
a mechanism able to include an energy loss during the impact. As shown in Ref.~\cite{prl_2019}, a possible route is to require that after each collision with the surface
the $x$ component of the particle's momentum is completely uncorrelated.
Under this assumption, the static source term vanishes and the equilibrium 
dynamical correlations can be evaluated analytically~\footnote{The 
details can be found in the SM of Ref.~\cite{prl_2019}.} and Eq.~(\ref{gold}) reduces to
\begin{equation}
\int_{0}^{+\infty}\ud z' \, \partial_{z'} u^x(z') \rme^{-z'^2+2\zeta z'}=
\frac{\tau k_{\mathrm{B}}}{2m} \partial_x T,
\label{eq:gold_gas}
\end{equation}
where $\zeta=z\sqrt{m\beta/2\tau^2}$ and a finite relaxation time $\tau$ 
has been introduced, to mimic the behaviour of an almost ideal gas, where some collisions appear.
The same hypotheses must be applied for the evaluation of the contributions arising 
from the dynamical correlations in Eq.~(\ref{massj}) and the final result reads:
\begin{align}
\big<\hat j^x(z)\big> &= \frac{\eta}{2}\frac{\partial_x T}{T} + \frac{\eta}{4}
\Biggl\{\mathrm{erf} \left(\sqrt{\frac{3}{2}}\frac{z}{\ell_g}\right) \notag \\
& \quad -\sqrt{\frac{3}{2\pi}}\,\frac{z}{\ell_g}\, \mathrm{Ei}\left[-\frac{3}{2}\left(\frac{z}{{\ell}_g}\right)^2\right]
\Biggr\} \frac{\partial_x T}{T},
\label{eq:massj_gold}
\end{align}
where $\ell_g\,=\,\tau \sqrt{2/(m\beta)}$ and $\mathrm{Ei}(\cdot)$ is the exponential integral function.
As already discussed, the solution of Eq.~(\ref{eq:gold_gas}) can be only found up to 
an additive constant, that has been fixed by imposing \hbox{no-slip} boundary 
on the mass current $\big<\hat j^x(0)\big>$, in accordance with the hypothesis 
introduced above.
Far from the surface, in the limit $z\gg\ell_g$, the last contribution 
in~(\ref{eq:massj_gold}) vanishes and the slip velocity reduces to the 
Maxwell's prediction:
\begin{equation}
v_{\infty} =\lim_{z\to\infty}\frac{\big<\hat j^x(z)\big>}{\rho_0} = \frac{3}{4}\,\frac{\eta}{\rho}\,\frac{\partial_x T}{T}.
\label{eq:maxwell_v}
\end{equation}

\subsection{Closed Channel}

A different configuration of the same model can be dubbed ``closed channel". In this case we imagine to place two hard walls, confining the system in the $x$ direction. A rigorous study of this problem is complicated by the 
absence of translational invariance which induces a dependence on $x$ of the external fields. However, in a sufficiently 
long channel and far enough from the two additional walls, we can assume that the translational invariance is recovered and the 
main effect of the presence of the additional boundaries is the constraint of vanishing of the integrated mass flow:
\begin{equation}
\int_0^h \ud z \,\bla \hat j^x(z)\bra =  0.
\label{j0}
\end{equation}
The fluid cannot flow freely along the channel, but a backflow must set in. This backflow is driven by a pressure gradient 
self-consistently generated by the \hbox{thermo-osmotic} flow close to the confining surfaces. In this respect, 
the closed channel configuration is particularly interesting from an experimental point of view because 
the bulk pressure gradient is in principle a measurable quantity~\cite{shukla_review_1984,barragankjelstrup_2016review} and provides quantitative
information on the thermo-osmotic phenomenon. In turn, the pressure gradient is 
determined by the choice of the unknown parameter $\gamma$: We already know that if $\gamma$ is given by Eq.~(\ref{open}) 
the pressure is uniform along the channel far from the walls. 
Every deviation from that value leads to a pressure gradient and then to a backflow. 
In order to find $\gamma$ we must start from the condition~(\ref{j0}) together with the definition~(\ref{massj})
which require that the quantity 
\begin{equation}
\int_0^t \ud t' \int_0^h \ud z \int \ud \br^\prime \bla \hat j^x(\br,t')\hat J_Q^x(\br^\prime) \bra_0 
\end{equation}
does not diverge.  Due to the translational invariance in the $(x,y)$ plane, this integral is just a function of $t$ 
which has to tend to infinity (see Eq.~(\ref{steady})). 
For a generic $\gamma$ this quantity is expected to diverge either to plus 
or to minus infinity as $t \to\infty$. 
The reason is the presence of the long, \hbox{non-integrable}, tails affecting the dynamical correlation functions of 
conserved currents~\cite{simpleliquids_fourth}. In particular both the \hbox{space-time} integrals of $\bla \hat j^x(\br,t')\hat J_H^x(\br^\prime) \bra_0$ 
and $\bla \hat j^x(\br,t')\hat j^x(\br^\prime) \bra_0$ are expected to diverge. Only a specific linear combination of these
two functions has a finite limit for $\tau\to\infty$. The coefficient of such a unique linear combination precisely identifies the correct 
choice for $\gamma$. In the adopted geometry, this ``magic" value will be state dependent but also will be a function of the channel 
width $h$: $\gamma(h)$. Equation~(\ref{gold}) together with the constraint~(\ref{j0}) and the explicit expression 
for the average mass current~(\ref{massj}) fully determine the velocity field $u^x(z)$ and the mass flux $\bla\hat j^x(z)\bra$.

Often the nano-channels confining the fluid are significantly wider than the fluid correlation
length and therefore it is important to ascertain the behavior of both the velocity profile and the ensuing pressure gradient 
in this limit. From the previous, exact, equations, it is possible to gain some general information on the large $h$ behavior.
Recalling that $\gamma(h)$, for a wide channel, must be close to its limiting value for $h\to\infty$: $\gamma(h) = h_m +\delta\gamma(h)$,
we can expand the static source term~(\ref{ss2}) at large $h$ and for $z$ far from the surface as 
\begin{eqnarray}
{\cal S}_s(z) &\sim& {\cal S}^\infty_s(z) -\kb T\,\rho_0\,\delta\gamma(h)\,\left (\frac{h}{2}-z\right )\nonumber \\
&\sim& \kb T\,\rho_0\,\delta\gamma(h)\,\left (z-\frac{h}{2}\right )
\end{eqnarray}
because the source term ${\cal S}^\infty_s(z)$ vanishes a few correlation lengths away from the confining surface. 
Therefore, in the central region of a wide channel, the gradient of the velocity field will be given by 
\begin{equation}
\partial_z u^x(z) \sim \frac{\rho_0}{\beta\eta}\,\delta\gamma(h)\,\left (z-\frac{h}{2}\right ) \,\partial_x\beta,
\end{equation}
where we assumed that the velocity field is a slowly varying function of $z$ on the range of the integral kernel ${\cal K}$.
The mass flux is easily evaluated from Eq.~(\ref{massj}) and, far from the walls becomes
\begin{eqnarray}
\langle \hat j(z)\rangle &\sim& \rho_0 u^x(z)\nonumber \\
&\sim& \rho_0\,\left [ \omega - 
\frac{\rho_0}{2\beta\eta}\,\delta\gamma(h)\,z\,(h-z) \,\partial_x\beta\right ],
\nonumber 
\end{eqnarray}
where $\omega$ is an integration constant. This expression provides the mass flux far from the surfaces. As previously 
noted, we expect that the flow close to the confining surfaces is not deeply affected by the width of the channel. Therefore, 
the vanishing of the integrated current can be written as 
\begin{equation}
0 = \int_0^h \ud z \,\bla\hat j(z)\bra \sim \rho_0\omega\,h - \frac{\rho_0^2\,h^3}{12\beta\eta}\,\delta\gamma(h)\,\partial_x\beta
+{\rm const}, 
\nonumber 
\end{equation}
where the additive constant accounts for the contribution of the thermo-osmotic flow close to the walls. 
This equation allows to determine the integration constant $\omega$ for large $h$:
\begin{equation}
\omega \sim \frac{\rho_0\,h^2}{12\beta\eta}\,\delta\gamma(h)\,\partial_x\beta + O\left(\textstyle{\frac{1}{h}}\right).
\end{equation}
Evaluating the fluid velocity profile, we finally obtain 
\begin{eqnarray}
v^x(z) &=& \frac{\langle\hat j(z)\rangle}{\rho_0} \nonumber \\
&\sim&  \frac{\rho_0\,h^2\,\delta\gamma(h)}{12\beta\eta}\,\left [ 1 - \frac{6\,z\,(h-z)}{h^2}\right ]\,\partial_x\beta\nonumber\\
&=& v^x(0)\, \left [ 1 - \frac{6\,z\,(h-z)}{h^2}\right ]
\label{vprof}
\end{eqnarray}
to leading order in $h$. 
The pressure gradient in the bulk region is obtained from the relation
\begin{eqnarray}
\partial_x p_0 &=& \big[ \partial_\beta p_0 + \gamma(h)\,\partial_{\beta\mu} p_0 \big] \,\partial_x\beta \nonumber \\
&=& \frac{\rho_0}{\beta}\,\delta\gamma(h)\,\partial_x\beta.
\label{drop}
\end{eqnarray}
Substituting Eq.~(\ref{drop}) into Eq.~(\ref{vprof}) we obtain a result consistent with macroscopic hydrodynamics:
\begin{equation}
v^x(z)=\frac{\partial_x p_0}{12\eta}\,\big [ h^2 - 6\,z\,(h-z)\big].
\label{vprof2}
\end{equation}
The slip velocity, i.e. the velocity of the bulk flow extrapolated at the wall is then related to the pressure gradient by $v^x(0)=\frac{h^2\,\partial_x p_0}{12\eta}$.
The velocity profile~(\ref{vprof2}) has indeed the typical Poiseuille form, expected from the \hbox{Navier-Stokes} equation, showing that the microscopic linear response formalism correctly reduces to the macroscopic 
approaches in the appropriate limits. Eqs.~(\ref{vprof}) and~(\ref{vprof2}) also suggest that 
the fluid velocity in the middle of the channel $z=\frac{h}{2}$ is finite and \hbox{non-zero} 
in the $h\to\infty$ limit only if the asymptotic scaling 
\begin{equation*}
\delta\gamma(h)\sim \partial_x p_0\sim \frac{1}{h^{2}}
\end{equation*}
holds, i.e., if the pressure gradient in a wide closed channel scales as the inverse square of its width. In this case also the slip velocity $v^x(0)$
attains a finite limit.

\section{Simulations}
\label{sim}

\begin{figure}

	\input{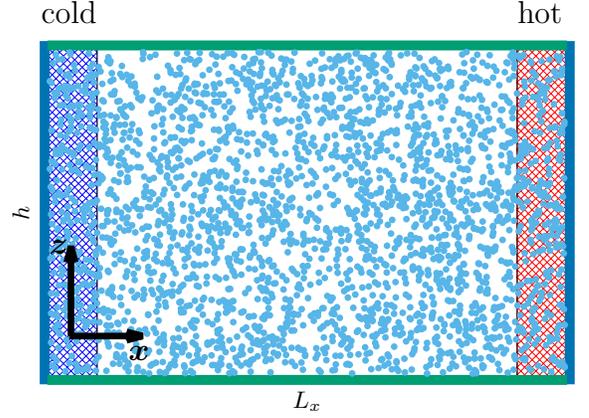}
        \caption{\label{fig:sys}Layout of a typical 2d simulation cell.
                 The two shaded regions are kept at constant temperature (cold and hot).
		 The confinement along the $x$ direction is guaranteed by reflective walls, whereas the
		 surfaces at fixed $z$, which induce the \hbox{thermo-osmotic} effects, are hard walls plus the finite range repulsive potential $V(z)$ defined in Eq.~(\ref{eq:Vr}).}
\end{figure}

In order to test the predicted relation between the pressure gradient and the fluid velocity in a closed channel~(\ref{drop}),  
we performed nonequilibrium molecular dynamics simulations in the \hbox{two-dimensional} geometry sketched in Fig.~\ref{fig:sys}. 
by use of the LAMMPS package~\cite{LAMMPS} (\texttt{http://lammps.sandia.gov}).
Particles interact through a pair potential of the \hbox{Lennard-Jones} form:
\begin{equation}
v(r) =
    \begin{cases}
            v_{\mathrm{LJ}}(r) - v_{\mathrm{LJ}}(r_c) & r \leq r_c\\
            0 & r > r_c
    \end{cases},
    \label{eq:ffpot}
\end{equation}
where the expression of the 12/6 LJ potential $v_{\mathrm{LJ}}$ reads
\begin{equation}
v_{\mathrm{LJ}}(r) =4\epsilon \left[ \left( \frac{\sigma}{r} \right)^{12} - \left( \frac{\sigma}{r} \right)^6 \right].
\label{eq:LJ126}
\end{equation}
The parameters $\epsilon$ and $\sigma$ represent the depth of the potential well
and the particle diameter respectively, whereas the
cutoff radius $r_c$ is set to $4.5\sigma$. The dimensional constants $\sigma$ and $\epsilon$, together with the
particle mass $m$, allow to define the standard time unit $\tau=\sigma\,\sqrt{\frac{m}{\epsilon}}$.
Two identical confining walls are set at $z=0$ and $z=h$. They are hard walls plus a finite-range repulsive potential $V(z)$ of the form: 
\begin{equation}
        V(z) =
    \begin{cases}
            k(z-z_0)^2 & z \leq z_0\\
            0 & z > z_0
    \end{cases},
    \label{eq:Vr}
\end{equation}
where $k = 0.1 \epsilon/ \sigma^{2}$ and $z_0 = 5\sigma$. 
The system is enclosed by two reflective walls, placed at $x=0$ and $x=L_x$, being $L_x$ the length of the channel.
Simulations are characterized by a time step $\delta t = 0.005\tau$. 
All systems undergo a first equilibration phase of $10^7$ time steps in order to reach 
a uniform temperature of $T=0.9\epsilon/k_{\text{B}}$ through a canonical sampling thermostat that uses global velocity
rescaling with Hamiltonian dynamics~\cite{par_bus} and a NVE time integration of the equation of motion.
Then, the thermal gradient is set in the $x$ direction controlling only the temperatures 
of the two thermostated regions highlighted in Fig.~\ref{fig:sys}.
During this stage, a constant temperature gradient $\partial_x T=0.0005 \epsilon/\kb \sigma$ develops
in the system, while the average temperature is kept at $T=0.9\epsilon/k_{\text{B}}$.
This transition phase lasts $6\times10^7$ time steps.
The last phase is the production one, where the previous temperature conditions are maintained 
and the desired properties are measured. This stage lasts up to $8\times10^8$ time steps.

\begin{figure}
	\input{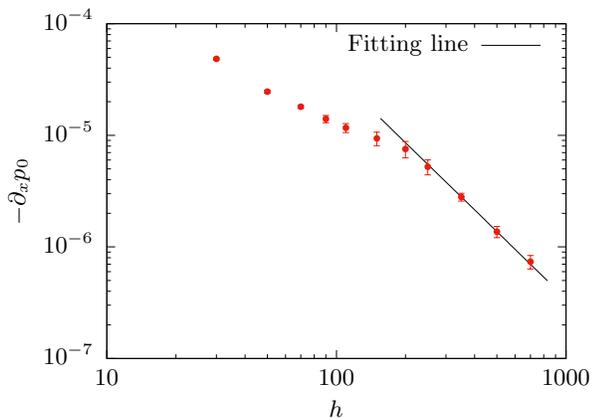}
	\caption{\label{fig:2}Scaling behavior of the bulk pressure gradient with the width of the channel. 
The black line represents the $h^{-2}$ asymptotic behavior.
Data refer to channels characterized by the same length $L_x=200$ and 
show $-\partial_x p_0$, being the bulk pressure gradient, negative in these systems. Length in units of $\sigma$, pressure gradient in units of $\epsilon/\sigma^{3}$}
\end{figure}

Two sets of systems characterized by a bulk density of $\rho_b \approx 0.54\sigma^{-2}$ were simulated.
In the first one, the length of the channel is kept fixed at $L_x=200\sigma$ while different widths are considered, 
ranging from $h=30\sigma$ up to $h=700\sigma$, with $N=2760$ and $N=74615$ particles respectively.
This series of simulations allowed us to verify the behavior of the bulk pressure gradient with the width of the channel:
As shown in Fig.~\ref{fig:2} numerical simulations confirm the expected 
$\partial_x p_0\sim h^{-2}$ behavior, represented by the black line, for wide systems.

\begin{figure}
       \input{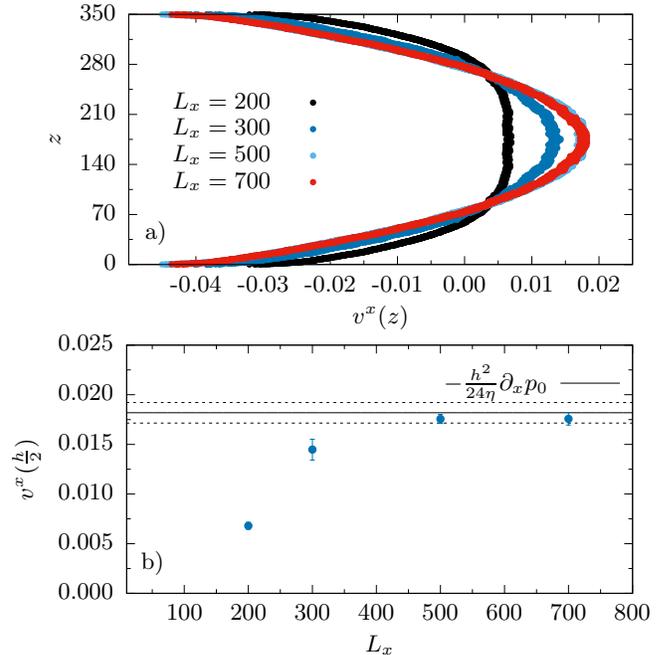}
        \caption{\label{fig:4vlength}a$)$: Velocity profiles for channels of height $h=350$ and different lengths $L_x$, from left to right $L_x=200$, $300$, $500$, $700$ (the last two curves are almost superimposed). 
b$)$: Comparison between the prediction of Eq.~(\ref{vprof2}) and the simulated data.
Lenght in units of $\sigma$, velocity in units of $\sigma / \tau$.}
\end{figure}

Equation~(\ref{vprof2}) provides a simple link between the bulk pressure gradient and the velocity in the middle of the channel.
This relation has been theoretically derived in the limit of very long channels $L_x\to\infty$, supporting the usual Poiseuille flow
profile. To verify this result, we performed a second set of simulations.
We considered four channels with the same width $h = 350\sigma$ (large enough to guarantee the 
asymptotic behavior of the bulk pressure gradient with $h$), but characterized by different lengths:
$L_x=200\sigma,\,300\sigma,\,500\sigma$ and $700\sigma$, with a number of particles ranging from $N=37179$ up to $N=130060$.
The need of simulating four different length values is clear if we look at Figure~\ref{fig:4vlength}, panel a), where the resulting velocity profiles are shown.
Increasing $L_x$ the shapes of the profiles change up to $L_x=500\sigma$, where the bulk parabolic behavior is recovered and the velocity profiles become independent on the channel length.
Shorter channels clearly induce more complex hydrodynamic patterns, violating our central assumption stating that the 
sole effect of the presence of the walls limiting the flow in the $x$ 
direction is the vanishing of the integrated current Eq.~(\ref{j0}). 
The velocity profile in the longest channels can be considered equal within simulation errors and a parabolic fit 
allows to measure the viscosity coefficient which turns out to be 
$\eta = 0.785\pm0.044\tau\epsilon/\sigma^{2}$.
Inserting this value in Eq.~(\ref{vprof2}) we can test the relation between the pressure 
gradient and the velocity at the center of the channel 
$v^x\left(\textstyle{\frac{h}{2}}\right)$. Panel b) of Fig.~\ref{fig:4vlength} indeed shows a remarkable agreement. 
It is important to note that the bulk pressure gradients obtained in the latter set of simulations are statistically equivalent.
This suggests that the channel length does not affect this observable and the results shown in Fig.~\ref{fig:2} maintain their validity.

\section{Conclusions}

The microscopic theory presented in Section~\ref{micro} allows to fully specify the properties of the nonequilibrium steady state of a 
confined fluid at \hbox{non-uniform} temperature in terms of the \hbox{fluid-fluid} and \hbox{fluid-walls} interactions. Linear response theory provides explicit 
expressions of all the quantities of interest, like mass or heat current, in terms of the static and dynamic structural properties of 
the fluid at equilibrium, which can be evaluated by use of liquid state theory or numerical simulations. 

The \hbox{take-away} messages emerging from 
this approach are $i)$ the key role played by the \hbox{fluid-wall} interface in driving the effect; $ii)$ the existence of two physically 
different mechanisms: A ``{\it static}" one related to the change of the equilibrium properties of the fluid near the confining surface and 
a ``{\it dynamic}" one, originated by the presence of momentum and energy transfer between the fluid particles and the walls during collisions.
The static mechanism gives rise to a flow within a few correlation lengths from the wall, or within the range of the \hbox{wall-particle} interaction.
The dynamic mechanism develops on the typical \hbox{length-scale} of the mean free path, which can be extremely large in diluted systems, and 
is likely to become the dominant effect in gases. The static and dynamic mechanisms parallel two phenomenological approaches developed respectively in 
liquids by Derjaguin, in the context of nonequilibrium thermodynamics, and in gases by Maxwell, in the framework of kinetic theory. It is reassuring 
that a first principle microscopic theory recovers these classical results in the appropriate limits. 
Numerical simulations~\cite{md_simu} showed that 
the extent of \hbox{thermo-osmosis} is particularly sensitive to the form of the \hbox{fluid-wall} interaction, which sets both the sign and the  
amount of the mass flow. 

Few predictions of the microscopic theory have been verified by numerical simulations in a simple two-dimensional slab geometry in 
Section~\ref{sim}, 
mimicking possible experimental realizations: Extrapolating the simulation data to a three dimensional system, 
we can estimate that confining a molecular liquid in a nanochannel  of radius $R\sim 1$ $\mu$m and imposing a temperature difference $\Delta T$ 
between the ends of the channel
gives rise to a pressure difference of the order of $\frac{\Delta P}{\Delta T}\sim  10^2 $ Pa/K.  

Thermo-osmosis is an interesting effect {\it per se}, being the simplest example of thermal force, 
and plays a relevant role in different physical frameworks, from engineering to biophysics, 
involving \hbox{temperature-driven} fluid flows through membranes. However, the most important role played by 
the \hbox{thermo-osmotic} mechanism probably occurs at the surface of colloidal particles immersed in a liquid or a gas,
where the ensuing fluid flow pushes the colloidal particles through the fluid, giving rise to thermophoresis. 

A microscopic study of \hbox{thermo-osmosis} is also instrumental for defining the correct boundary conditions for effective macroscopic 
approaches, based on hydrodynamics and the \hbox{Navier-Stokes} equation, describing fluid flow in confined systems. Understanding 
what happens in a fluid within a few molecular diameters from the boundary surfaces
allows to quantitatively specify the slip induced by the presence of temperature gradients at the confining walls. 

 Finally, our results suggest that numerical simulations of liquids in narrow pores under thermal gradients may be efficiently performed by first evaluating the static source term {\it via} equilibrium simulations, which provide the effective force driving the flow. Then, in a further nonequilibrium simulation at uniform temperature, the effective force previously found can be used to mimic the effects of the thermal gradient. This procedure,
pioneered in Refs.~\cite{ganti17,ganti18}, is now substantiated by Linear Response Theory.
\vspace{1cm}

\section{Appendix}
\label{sec:appendix}
\subsection{Static source term in planar geometry}
\label{sec:sst_divergence}

\subsubsection*{Derivation of Eq.~(\ref{source1})} 
The starting point of this derivation is the evaluation of the derivative of Eq.~(\ref{ss})
\begin{align}
\partial_z{\cal S}_s(z) &=
\partial_\beta p_0^{xx}(z)+ \gamma \, \partial_{\beta\mu}p_0^{xx}(z) \notag \\
& \qquad-\int \ud\br^\prime \,x^\prime\,\left \langle \partial_z\hat J_j^{xz}(\br) \,
\hat {\scr P}(\br^\prime) \right \rangle_0. \label{der_Sz_1}
\end{align}
The first two contributions at r.h.s.~can be evaluated from the definition of the pressure tensor at equilibrium
\begin{equation}
p_0^{\alpha\beta}(z)=\Big \langle \hat J_j^{\alpha\beta}(\br)\Big \rangle_0 ={{\cal Q}_0}^{-1}\int \ud \Omega  \, J_j^{\alpha\beta}(\br)\,e^{-\beta\hat H-\beta\mu \hat N},\notag
\end{equation}
where ${\cal Q}_0$ is the partition function of the grand canonical distribution function $\exp (-\beta\hat H-\beta\mu \hat N)$, and read
\begin{align}
\partial_\beta p_0^{xx}(z)&=-\int \ud\br^\prime \,\left \langle \hat J_j^{xx}(\br) \left[ \hat {\cal H}(r^\prime) -\big \langle {\hat{\cal H}}(r^\prime)\big \rangle_0 \right] \right\rangle  \notag \\
\partial_{\beta\mu} p_0^{xx}(z)&= \int \ud\br^\prime \,\left \langle \hat J_j^{xx}(\br) \big[ \hat \rho(r^\prime) -\big \langle \hat\rho(r^\prime)\big \rangle_0 \big] \right\rangle \notag
\end{align}
By defining 
\begin{equation}
\Delta\hat{\scr P}(\br^\prime) = \hat{\scr P}(\br^\prime) - \big<\hat{\scr P}(\br^\prime) \big>_0\notag
\end{equation}
with $\hat{\scr P}$ given by~(\ref{p}), 
we can write the sum of the first two contributions as
\begin{equation}
\big[\partial_\beta + \gamma \, \partial_{\beta\mu}\big]p_0^{xx}(z)=-\int \ud\br^\prime \left \langle \hat J_j^{xx}(\br)\,\Delta\hat{\scr P}(\br^\prime)\right  \rangle_0.
\notag
\end{equation}
In addition, the translation invariance along the $x$ direction implies
\begin{align}
\int \ud\br^\prime \, x^\prime& \left\langle \partial_z \hat J_j^{xz}(\br)
\bla\hat{\scr P}(\br^\prime)\bra_0 \right\rangle_0 \notag\\
&=\bla  \partial_z\, \hat J_j^{xz}(\br) \bra \int \ud\br^\prime \, x^\prime \bla\hat{\scr P}(\br^\prime)\bra_0=0, \notag
\end{align}
and Eq.~(\ref{der_Sz_1}) can be finally written as
\begin{equation}
\partial_z\,{\cal S}_s(z) = -\int \ud\br^\prime \left \langle \Big [ \hat J_j^{xx}(\br)+x^\prime\, \partial_z \hat J_j^{xz}(\br)\Big ]
\,\Delta\hat{\scr P}(\br^\prime)\right  \rangle_0.  \nonumber 
\end{equation}
Next we recognize that the translational invariance in the $(x,y)$ plane forces the averages in the previous equation 
to depend only on $(x-x^\prime)$ and $(y-y^\prime)$, proving the identity
\begin{widetext}
\begin{align}
\int \ud\br^\prime \,x^\prime \,\partial_\alpha\,\big<\hat J_j^{x\alpha}(\br)\,\Delta\hat{\scr P}(\br^\prime)\big>_0 =&
\int \ud\br^\prime \,\left [ 
x^\prime \,\partial_z\,\big<\hat J_j^{xz}(\br)\,\Delta\hat{\scr P}(\br^\prime)\big>_0 +
\big<\hat J_j^{xx}(\br)\,\Delta\hat{\scr P}(\br^\prime)\big>_0 \right ] \notag \\
&- \int \ud\br^\prime \, \left [ \partial_{x^\prime} \,x^\prime
\left\langle \hat J_j^{xx}(\br) \,\Delta\hat{\scr P}(\br^\prime)\right\rangle_0 
+ \partial_{y^\prime}\,x^\prime\,\left\langle \hat J_j^{xy}(\br)
\,\Delta\hat{\scr P}(\br^\prime)\right\rangle_0 \right ]. 
\notag 
\end{align}
\end{widetext}
The last line is a total divergence which vanishes upon integration if the static correlation function
decays sufficiently fast to infinity, while the first contribution is precisely minus the derivative of the static source term:
\begin{equation}
\partial_z\,{\cal S}_s(z) = -\int \ud\br^\prime \,x^\prime \,\partial_\alpha
\bla\hat J_j^{x\alpha}(\br)\,\Delta\hat{\scr P}(\br^\prime)\bra_0.
\nonumber
\end{equation}
\subsubsection*{Derivation of Eq.~(\ref{final})}
In the case of a \hbox{$z$-dependent} external potential, the continuity equation 
for the $x$ component of the momentum density operator~(\ref{momc}) at $t=0$ reads
\begin{equation}
\frac{\ud j^x(\br)}{\ud t} + \partial_\alpha \hat J_j^{x\alpha}(\br) = 0.
\notag
\end{equation}
This Equation can be used to substitute the divergence term by minus the time
derivative of the current. However, a general property of the time-dependent correlation functions at equilibrium allows to
move the time derivative to the second operator:
\begin{equation}
\partial_z\,{\cal S}_s(z) = -\int \ud\br^\prime \,x^\prime \,
\bigg\langle\hat j^{x}(\br)\,\frac{\ud\Delta\hat{\scr P}}{\ud t} (\br^\prime)\bigg\rangle_0,
\end{equation}
where the derivative is again evaluated at $t=0$. Finally, we recall that also the operator $\hat{\scr P}(\br)$ satisfies a
continuity equation
\begin{equation}
\frac{\ud \hat{\scr P}(\br)}{\ud t} + \partial_\alpha \hat J_Q^{\alpha}(\br) = 0 
\end{equation}
in terms of the heat current~(\ref{heat}). Substituting and integrating by parts we finally get:
\begin{equation}
\partial_z\,{\cal S}_s(z) = -
\int \ud\br^\prime \, \bla\hat j^{x}(\br)\,\hat J_Q^{x} (\br^\prime)\bra_0.
\nonumber
\end{equation}
If we substitute in this equation the microscopic expressions of the momentum~(\ref{eq:mic_mom_curr}) and heat~(\ref{heat}) current 
\begin{align*}
\hat j^{x}(\br) &= 
\sum_i p_i^x \,\delta(\br - \bq_i), \\
\hat J^x_Q(\br)&=\sum_i \left[ \frac{\hat h_i}{m} -\gamma\right]p_i^x \,\delta(\br - \bq_i)+\sum_i \Gamma_i^{x\nu}(\br)\frac{p_i^\nu}{m}
\end{align*}
and we evaluate analytically the equilibrium average, the integrated 
correlation function becomes independent of the specific choice of the integration path in Eq.~(\ref{prob}) and 
is expressed in terms of the {\it virial transverse} enthalpy density in thermal equilibrium:
\begin{align}
\quad &h_0^v(z)=\left [ \frac{5}{2}\kb T +V(z)\right ] n_0(z) \nonumber  \\
&+\frac{1}{2} \int \ud\br^\prime \, n_2(\br,\br^\prime)
\left [ v(s) -  \frac{\ud v(s)}{\ud s} \,\frac{(x-x^\prime)^2}{s}\right ]_{s=|\br-\br^\prime|}, \nonumber 
\end{align}
where $n_0(z)$ is the average equilibrium particle density $m\,n_0(z) = \rho_0(z)$  and $n_2(\br,\br^\prime)$ 
is the \hbox{two-particle} equilibrium static correlation function.

\subsection{Cylindrical geometry}
\label{pore}

\begin{figure}
\centering
\begin{tikzpicture}[scale=0.9,every node/.style={scale=0.9}]
  \def\L{3.0} 
  \def\R{1.2}
  \def\dr{0.1}
  \def\a{3.0}
  \def\ang{8}
  \def\angp{20}
  \def\angdr{14}
  \coordinate (O) at (0,0);
  \coordinate (R) at (\ang:\R);
  \draw[thick] (\angp:\L) --++ (\angp:1.8*\R);
  \draw[border,
    top color=slab!50,bottom color=slab!50,middle color=slab!20]
    (\angp+90:\R+4*\dr) --++ (\angp:\L) arc(\angp+90:\angp-90:\R+4*\dr) --++ (\angp-180:\L) arc(\angp-90:\angp-270:\R+4*\dr);
  \draw[border,fill=liquid] (O) circle(\R);
  \draw[border] (O) circle(\R+4*\dr);
  \draw[->,thick] (\angp:\R-0.01) -- (\angp-180:1.7*\R) node[left] {$z$};
  \draw[->,thick] (O) -- (+85:2*\R) node[left=8,below=1] {${r}$};
  \draw[-] (O) -- (-40:\R) node[midway,left=10,below=-3] {{$R$}};
\end{tikzpicture}
\caption{Schematic representation of the cylindrical geometry.}
\label{fig:geocyl}

\end{figure}
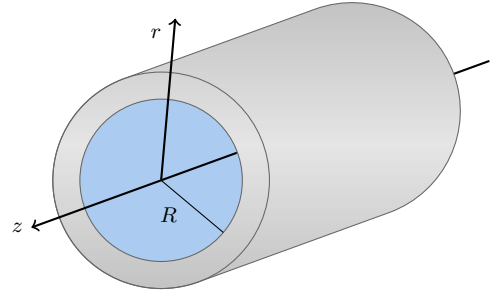

Here we report the explicit expression of the Eqs.~\hbox{(\ref{gold}--\ref{sd})}  in 
the case of cylindrical geometry, appropriate for a pore or a nanotube (see Fig.~\ref{fig:geocyl}). 
The derivation closely parallels the 
analysis performed for a fluid in a slab. The role of the coordinate $z$ is now played by the radial coordinate $r$ which 
varies between $0$, at the center of the tube, and $R$ where the confining surface is placed. Our Ansatz for the solution 
is the natural generalization of Eqs.~(\ref{beta}--\ref{u}), the main difference being the formal expression 
of the divergence in cylindrical coordinates. We just quote the equations replacing Eqs.~(\ref{gold}--\ref{sd}):
\begin{equation}
\notag
\int_0^{R} \ud r^\prime r^\prime \,{\cal K}(r,r^\prime)\,  \partial_{r^\prime}u^x(r^\prime) = \partial_x\beta\,\left [ {\cal S}_s(r)+{\cal S}_d(r)\right ],
\end{equation}
where the kernel ${\cal K}(r,r^\prime)$ is now given by 
\begin{equation}
{\cal K}(r,r^\prime) = \beta \int \ud x^\prime \int\ud\phi^\prime  
\int_0^\infty \ud t^\prime  \bla\hat J_j^{xr}(\br,t^\prime)\hat J_j^{xr}(\br^\prime)\bra_0,  
\nonumber
\end{equation}
while the static and dynamic source terms are 
\begin{eqnarray}
{\cal S}_s(r) &=&
-\frac{\kb T}{r}\,\int_0^r \ud r^\prime \, r^\prime \big [ h(r^\prime) - \gamma\,\rho_0(r^\prime)\big]; 
\notag
\\
{\cal S}_d(r) &=&
\int_0^\infty \ud t \int \ud\br^\prime \bla \hat J_j^{xr}(\br,t)
\hat J^x_Q(\br^\prime)\bra_0.
\notag
\end{eqnarray}

\bibliographystyle{apsrev4-2}
%

\end{document}